# Electron-proton transfer mechanism of excited state hydrogen transfer in phenol–$(NH_3)_n$ ($n$ = 3 and 5)

Mitsuhiko Miyazaki,[a] Ryuhei Ohara,[a] Claude Dedonder,[b] Christophe Jouvet,*[b] and Masaaki Fujii*[a]

[a]Laboratory for Chemistry and Life Science, Tokyo Institute of Technology, 4259-R1-15, Nagatsuta-cho, Midori-ku, Yokohama 226-8503, Japan.
[b]CNRS, Aix Marseille Université, Physique des Interactions Ioniques et Moleculaires (PIIM) UMR 7345, 13397 Marseille cedex, France

Corresponding authors: CJ (christophe.jouvet@univ-amu.fr) and MF (mfujii@res.titech.ac.jp)

**Abstract**

Excited state hydrogen transfer (ESHT) is responsible to various photochemical processes of aromatics including photoprotection of nuclear basis. Its mechanism is explained by the internal conversion from aromatic ππ* to πσ* states via conical intersection. It means that the electron is transferred to a diffuse Rydberg like σ* orbital apart from the proton migration. This picture means the electron and the proton are not move together and its dynamics are different in principle. Here, we have applied the picosecond time-resolved near infrared (NIR) and infrared (IR) spectroscopies to the phenol–$(NH_3)_5$ cluster, the bench mark system of ESHT, and monitored the electron transfer and proton motion independently. The electron transfer monitored by the NIR transition rises within 3 ps while the overall H transfer detected by the IR absorption of NH vibration appears with the lifetime of ≈20 ps. It clearly proves that the electron motion and proton migration are decoupled. Such the difference of the time-evolutions between the NIR absorption and the IR transition has not been detected in the cluster with three ammonia molecules. We will report full of our observation together with theoretical calculations of potential energy surfaces of ππ* and πσ* states, and will discuss the ESHT mechanism and its cluster size-dependence between $n$ = 3 and 5. It is suggested that the presence and absence of a barrier in the proton transfer coordinate cause the different dynamics.

## 1. Introduction

Proton and hydrogen atom transfer (PT and HT) are one of the simplest but important chemical reactions, which appear in various systems from simple solution chemistry to biological systems and even in fuel cell chemistry. In particular, PT/HT triggered by photoexcitation, excited state proton/hydrogen transfer (ESPT/ESHT) attracts strong attentions because of its controllability by photoexcitation.[1] In solution phase, PT studies are further expanded to proton-coupled electron transfer (PCET).[2] In gas phase spectroscopy, ESPT was a main issue to study in the solvated clusters of photoacids, such as phenol (PhOH),[3] naphthol[4] etc.[5] The size dependence of ESPT reactions are studied extensively in long time. PhOH–$(NH_3)_n$ clusters are also such clusters that were believed to have ESPT.[3, 6] At the last year of the 20th century, ESHT was proposed instead of ESPT in PhOH–$(NH_3)_n$ from the detection of the long-lived species after photoexcitation.[7] Soon, ESHT was established by various experimental evidences.[8] In particular, the spectroscopic evidence that the intermediates is •$H(NH_3)_n$ radical, not a protonated one, gives the strong support to the paradigm shift from ESPT to ESHT (see Figure 1).[8a-f] Now ESHT is a *de facto* standard of mechanism in gas phase photochemistry, and various photochemical mechanism are redefined by ESHT such as indole,[9] tryptophan aromatic amino acid cations,[10] and many other systems.[11]

In theory, the mechanism of ESHT is described as an internal conversion from an initially prepared aromatic ππ* excited state to a diffuse and repulsive πσ* state via a conical intersection (CI).[12] The ππ* state is usually the initially prepared state in the aromatic photochemistry. Theoretical calculations show that the potential energy surface (PES) of ππ* crosses the repulsive PES of πσ* when the O-H or N-H bond is extended.[12] At the crossing point, the two PESs are connected by CI and then the ππ* electronic state goes to πσ*. The πσ* PES has an additional crossing with $S_0$ PES, and the photoexcited molecule is efficiently quenched by the internal conversion to $S_0$. The short lifetime of aromatic chromophores including nucleic bases are interpreted as a result of the conversion to $S_0$ via ESHT.[12b, 13] For clusters, the internal conversion to $S_0$ is mostly negligible because solvent stabilization of πσ* PES accelerates the H atom motion enough to go over CI to $S_0$.[12] Then the X-H bond can be simply elongated, resulting in H atom transfer to the solvent moiety. In the case of the PhOH–$(NH_3)_n$ clusters, ESHT generates the neutral radical by the H atom transfer i.e. PhOH–$(NH_3)_n$ → PhO• + •$H(NH_3)_n$.

The radical product can be detected by laser spectroscopy coupled with mass spectrometry. Experimentally, H atom transfer in clusters provides a clear signature of ESHT, because the stable reaction product, such as •$H(NH_3)_n$, has a surplus electron in the 3s Rydberg orbital, which 1) gives a low ionization energy, 2) shows characteristic 3p-3s Rydberg transitions in the near infrared (NIR) region,[14] and 3) exhibits characteristic vibrational transitions with high intensity.[8d] These characteristics provided the first spectroscopic evidence for ESHT and the NIR and infrared (IR) spectroscopy of the products provided clear confirmation.[8c]

The hydrogen atom transfer sounds like a simple H atom migration in which a proton and 1s electron move together. However, the mechanism of ESHT is far from the words "hydrogen atom transfer". The σ* orbital that the electron occupies during ESHT is a Rydberg-like diffuse one, and the σ* occupation has charge transfer character because the diffuse σ* orbital deviates rather ammonia moiety.[12] When ESHT is completed, the σ* electron becomes the pure 3s-Rydberg electron of the radical product. The corresponding proton transfer from the OH group to the solvent moiety proceeds independently from the electron motion. Thus, the electron motion and proton transfer are essentially decoupled in ESHT (see Figure 1). It means that the electron-proton decoupling is the fundamental issue in the mechanism of ESHT, however it has not been revealed experimentally.

Here, we attempt to detect the electron-proton decoupling in ESHT by a time-resolved spectroscopy for the PhOH–$(NH_3)_n$ ($n$ = 3 and 5) clusters, which are prototypical systems of ESHT. The ESHT in PhOH–$(NH_3)_n$ has two characteristic points: 1) a new N-H bond formation which produces •$NH_4$ and 2) the electron occupation of the 3s-Rydberg orbital of the solvent moiety via the Rydberg-like σ* orbital. The former can be detected by IR spectroscopy by the characteristic and intense N-H stretching vibrational transitions of •$H(NH_3)_n$ in the 3 μm region.[8d] Thus, the appearance of the strong NH stretching vibrations indicates the completed transfer of the proton and the electron from PhOH to the ammonia moiety. The latter can be observed by the NIR absorption due to Rydberg-Rydberg transitions originating from the electron in the Rydberg-like σ* and/or 3s-Rydberg orbitals. In general, the Rydberg-Rydberg transitions are very strong and are easily distinguished from valence transitions. In actual, the oscillator strength of the 3p-3s and 3p-σ* transitions are calculated at ≈0.4 and ≈0.2 which is more than 200 times stronger than the valence transitions (see Supplementary Information of ref. 15[15]). Therefore the appearance of the intense NIR transition indicates the electron transfer to the Rydberg orbital located on the solvent moiety. Since both are essentially independent phenomena, the comparison of the time-evolution of the NIR and IR transitions gives evidence whether electron and proton transfers are decoupled or not.

$S_1$-$S_0$ electronic transitions of PhOH–$(NH_3)_n$ have been reported previously.[8c] For clusters with $n \leq 4$, the spectra show well-resolved structures, which suggest a slow ESHT reaction (24 ps for $n$ = 3).[8b, 8f] In contrast, PhOH–$(NH_3)_5$ gives only a broad absorption, which indicates significant change of reaction mechanism of ESHT.[8c] For this reason, we focus on the time-resolved measurements for the $n$ = 5 cluster. The time-resolved spectroscopy of $n$ = 5 has been reported partly in the letter paper.[15] Here, we will report full of our observation together with theoretical calculations of PESs of ππ* and πσ* states, and will discuss the ESHT mechanism and its cluster size-dependence between $n$ = 3 and 5.

## 2. Methods

### 2-1 Experimental methods

The picosecond time-resolved IR/NIR (ps-TRIR/ps-TRNIR) spectroscopy of PhOH–

$(NH_3)_5$ was performed utilizing time-resolved ultraviolet (UV) -IR/UV-NIR ion dip spectroscopy, of which the principle has been described elsewhere.[8b, 8f, 15] The excitation scheme is illustrated in Figure 2. Briefly, PhOH–$(NH_3)_5$ was excited to the $\pi\pi^*$ state by a picosecond UV pulse $\nu_{exc}$ (35570 $cm^{-1}$), and the ESHT reaction was triggered. The reaction product, •$H(NH_3)_5$, was ionized by a nanosecond UV laser pulse, $\nu_{ion}$ that was irradiated 200 ns after $\nu_{exc}$, and the population of the reaction product was monitored at a mass of $H^+(NH_3)_5$ using a time-of-flight mass spectrometer. A picosecond tunable IR laser, $\nu_{IR}$, was triggered $\Delta t$ picosecond after $\nu_{exc}$ and was scanned over the NH stretching vibration region. If $\nu_{IR}$ is resonant with vibrational levels of the transient species, the population of •$H(NH_3)_5$ is depleted by the vibrational predissociation of the transient species. Therefore, the vibrational transitions of the transient species $\Delta t$ after the UV excitation was measured as a decrease of the monitored ion current. To measure time evolution of an IR transition, the delay time $\Delta t$ was scanned using an optical delay stage while fixing the frequency of $\nu_{IR}$ to that of the transition. If a tunable picosecond NIR pulse, $\nu_{NIR}$ is introduced instead of $\nu_{IR}$, time evolution of the electronic transition can be recorded. The ps-TRNIR spectra were measured between 4500 $cm^{-1}$ and 9000 $cm^{-1}$. The signal was measured at every 500 $cm^{-1}$ step after careful check of the laser beam alignments because of the wide spectral range.

A schematic diagram of the experimental apparatus is shown in Figure S1. A femtosecond mode-locked Ti:sapphire laser was regeneratively amplified. The amplified 800 nm output was separated into three pulses to pump three independent optical parametric generator/amplifiers (TOPAS; Light Conversion). One of these pulses was frequency doubled and was used as $\nu_{exc}$. The idler output of the second OPA was used as $\nu_{NIR}$. The $\nu_{IR}$ pulse was obtained by difference frequency generation between the second harmonic of the idler output of the third OPA and a part of the fundamental pulse in a KTA crystal. The pulse width of the picosecond pulses was 3 ps while the energy resolution was 12 $cm^{-1}$. These pulses were combined coaxially by beam combiners and focused by a lens with $f$ = 300 mm. The third harmonic output of an $Nd^{3+}$:YAG laser was used as $\nu_{ion}$. $\nu_{ion}$ was introduced into the vacuum chamber in a counter propagating manner to the picosecond pulses, and was focused by a 300 mm focal length lens.

The PhOH vapor at room temperature was seeded in $NH_3$/Ne 0.1% premix gas, and the mixture was expanded into the vacuum chamber through a pulsed valve at a stagnation pressure of 2 bars. The jet expansion was skimmed by a skimmer with a diameter of 2 mm, and resultant molecular beam was introduced into the ionization region of the TOF spectrometer, where the clusters interacted with the laser pulses. The system was operated at 10 Hz. The ion current was amplified and integrated by a digital boxcar to store the signal in a personal computer.

**2-2 Theoretical methods**

Ground states structures have been taken from previous work.[8d] They were obtained at the MP2 level using 6-31++G(d,p) as implemented in Gaussian 98.[16] The excited states energies have been calculated at the CC2 level using the TURBOMOLE program package,[17] making use of the resolution-of-the-identity (RI) approximation for the evaluation of the electron-repulsion integrals.[18] Calculations were performed with the correlation-consistent polarized valence double-zeta (aug-cc-pVDZ)[19] basis set adding diffuse functions in order to calculate properly the Rydberg πσ* states. The accuracy of the method with this basis or cc-pVDZ set can be evaluated from the previous comparison between experimental value and calculated electronic transitions and is typically 0.15eV.[20] The potential curves are obtained by stretching the OH distance keeping the other coordinate fixed at the ground state geometry.

In order to simplify the discussion, we name the two lowest excited states as ππ* and πσ* although this nomenclature would require the $C_s$ symmetry of the molecular system, which is not the case. However, the main orbital involved in the excited states are still keeping a very strong π valence or diffuse Rydberg-like σ* character (see Figure 10). Thus one should then keep in mind that what we call the ππ* state which at long O-H distance is the excited state proton transfer state has some πσ* character, and vis-versa, the hydrogen transfer state πσ* has some ππ* character.

## 3. Results and Discussion

From the previous study by the nanosecond "static" spectroscopy, we have several information on the ESHT reaction mechanism in the PhOH–$(NH_3)_5$. The PhOH–$(NH_3)_5$ cluster has a bicyclic hydrogen-bonded structure in $S_0$ (Figure 3). Photoexcitation of the cluster triggers the O-H cleaving by ESHT, and the H atom transferred to the ammonia moiety. According to the analogy to the $n$ = 3 cluster,[8b, 8f] the H atom will be accepted by the ammonia molecule which directly H-bonded to the PhOH. This causes the selective generation of •$H(NH_3)_5$ in $C_{3v}$+1 geometry (memory effect, see Figure 3). The $C_{3v}$+1 product is a meta-stable product and thus it will isomerize to the most stable product in the $T_d$ symmetry which is 1.6 kcal/mol stable than the $C_{3v}$+1 product. The formation of the $T_d$ product is proved by the NIR spectrum observed at 180 ns after the excitation to $S_1$, which is shown in the bottom of Figure 4.[8c] The strong broad signal centered at 6000 $cm^{-1}$ is assigned to the 3p-3s Rydberg transitions of the $T_d$ product based on the theoretical calculations.[8c, 21] The valence transition from $S_1$ ππ* state is 2 orders of magnitude weaker than the Rydberg transition and cannot contribute the transition.[15] Since the $T_d$ product is the final reaction product, the ESHT reaction is completed until 180 ns.

To detect the transient species, we applied ps-TRNIR spectroscopy to PhOH–$(NH_3)_5$. The top three spectra in Figure 4 present the ps-TRNIR spectra of PhOH–$(NH_3)_5$ with the delay time of 1–200 ps. Ten times expanded spectra are also shown by the blue traces. The time-resolved spectra show that the 6000 $cm^{-1}$ transition is very weak at 1 ps, and grows according to the increase of the delay time. The signal in lower frequency than 7000 $cm^{-1}$ shows essentially the same time-evolution. On the other hand, the NIR signal above 8000 $cm^{-1}$ appears rapidly and becomes weaker when the delay time increases. Such different behavior for the delay time clearly shows the existence of a transient species which have absorption above 8000 $cm^{-1}$. The quantum chemical calculations suggest two possible transient species in this region. One is the first-formed ESHT product of $C_{3v}$+1 symmetry of which the $3p_{x,y,z}$-3s transitions are calculated in the range of 6700–7700 $cm^{-1}$.[8c] Its oscillator strength is almost a comparable to the final $T_d$ product.[15] Another is 4p Rydberg-σ* transition of the complex originated from the occupation of electron in the πσ* orbital, of which the transition energy is 0.85 eV (6900 $cm^{-1}$). The 3p-σ* transition of the complex is also expected in the spectrum at 0.49 eV (4000 $cm^{-1}$) but it may not be clear because of overlapping of the transition of the $T_d$ species.

The time-evolutions of the NIR signals at 6000 $cm^{-1}$, 7000 $cm^{-1}$ and 8000 $cm^{-1}$ are shown in Figure 5. The signal at 6000 $cm^{-1}$ which corresponds to the $T_d$ product grows gradually and becomes constant at around 100 ps. On the other hand, the signal at 8000 $cm^{-1}$ rises rapidly and decays slowly. The rise of the signal is the same as the response function of the lasers, thus we cannot determine the exact rise time. The ultrafast rise of the signal is consistent with the broad spectral feature of the UV absorption transition of PhOH–$(NH_3)_5$.[8c] The signal decays after the sharp rise and becomes almost constant after ≈100 ps. The constant signal after ≈100 ps is attributed to the signal of the $T_d$ species which has strong intensity at this delay. The time-evolution at 7000 $cm^{-1}$ shows the sharp rise and the signal is constant after the rise. It is explained by the overlapping of decreasing signal at 8000 $cm^{-1}$ and increasing signal of the $T_d$ species. From these results, we concluded that the observed ultrafast rise at 8000 $cm^{-1}$ corresponds to the 3p-3s transition of the $C_{3v}$+1 product and/or the Rydberg-σ* transition in the charge transferred complex. In both cases, this signal is signature of the electron transfer from the aromatic π* to the σ* orbitals. It should be emphasized that such transient species before the formation of the $T_d$ species can be detected only by the ps-TRNIR spectroscopy.

The transient species which show the electron transfer is detected by the ps-TRNIR spectroscopy. As we described in the introduction, the N-H bond formation dynamics can be observed independently by the IR spectroscopy in 3 μm region. Figure 6 shows the ps-TRIR spectra of PhOH–$(NH_3)_5$. The delay time is indicated besides the each spectrum. The theoretical IR spectra of both the $C_{3v}$+1 and $T_d$ products are also shown at the bottom of the Figure 6. Here, the $T_d$ product will show the intense vibrational transitions at 2965 $cm^{-1}$ (H-bonded NH stretching of •$NH_4$) and 3260 $cm^{-1}$ (free NH stretching of $NH_3$), while the $C_{3v}$+1 species is calculated to give the IR absorption at 3217 $cm^{-1}$ (free NH stretching of $NH_3$). Both are clearly different each other as shown in the theoretical spectra, thus time-resolved IR spectroscopy in this frequency region can reflect the time-evolution of the nascent ESHT product independently from the NIR spectroscopy. The ps-TRIR spectra show the strong transition at ≈3200 $cm^{-1}$ and a broad absorption at ≈3000 $cm^{-1}$. Both are gradually grow in intensity with increasing delay time, however no vibrational transition showing a sharp rise was found. To confirm the absence of the ultrafast rise of the signals, we fixed the IR laser frequency to the bands at 2983 $cm^{-1}$, 3165 $cm^{-1}$ and 3248 $cm^{-1}$ (indicated by the letters A, B and C in the spectra, respectively) and scanned the delay. The bands A and C can be assigned to H-bonded NH stretching vibration and of •$NH_4$ and the free NH stretching bands of $NH_3$ in the $T_d$ product, respectively. The signal at B should be sensitive to the presence of the nascent $C_{3v}$+1 product. However, the most important point is that none of time-evolutions shown in Figure 7 shows an ultrafast rise, and the rises can be fitted by single exponential functions with the lifetimes of ≈20 ps.

We applied the ps-TRNIR and ps-TRIR spectroscopy to monitor the electron transfer and the NH bond formation independently. The ps-TRIR spectroscopy in the NH stretching vibrational region detects only the time-evolution of ≈20 ps rise while the NIR absorption at 8000 $cm^{-1}$ shows the fast rise of the signal that corresponds to the electron occupation of the Rydberg like σ* and/or 3s Rydberg orbital that has a charge transfer (CT) character (see Figure 3). Since the ps-TRIR spectra do not give the signature of the $C_{3v}$+1 product, it would be natural to assume that the population of the $C_{3v}$+1 product is negligible in the dynamics. It can be confirmed by the rate equation analysis on the time-evolutions at 6000 $cm^{-1}$, 7000 $cm^{-1}$ and 8000 $cm^{-1}$ shown in Figure 5. The detail of the rate equation analysis is described in SI. Briefly, the formation of the charge transfer complex from $S_1$ is assumed to be much faster than 3 ps, and successive reactions, i.e. the proton transfer to the $C_{3v}$+1 product and its isomerization to the final $T_d$ product, are included with rate constants $k_{\mathrm{S_1-CT}}$, $k_{\mathrm{CT}-C_{3v}+1}$, and $k_{\mathrm{Isomerization}}$. The best fits were indicated by the broken black curves in the Figure 5 with $1/k_{\mathrm{S_1-CT}}$ << 3 ps, $1/k_{\mathrm{CT}-C_{3v}+1}$ = 20.2 ps, and $1/k_{\mathrm{Isomerization}}$ ($= 1/k_{C_{3v}+1-T_d}$) = 0.4 ps. The analysis well reproduced the observed time-evolution of the NIR transitions. From the fitting, the populations of each species are drawn in Figure 5B. The population of the $C_{3v}$+1 species is limited thus the absence of its vibrational signature can be reasonably understood. Thus, the results strongly indicate that chemical bond formation is not complete in the nascent ESHT product due to the electron transfer. This clearly demonstrates that the electron transfer is decoupled from the transfer of the proton in ESHT.

The ESHT reactions rate in PhOH–$(NH_3)_n$ ($n$ = 1 and 3) are 1.1 ns and 24 ps.[8b, 8f, 22] In this work, the first step of ESHT in PhOH–$(NH_3)_5$ is measured to be faster than 3 ps, and is completed with ≈20 ps including the isomerization to the most stable $T_d$ products. This is clearly faster than that in the smaller cluster. For the $n$ = 3 cluster, the time-evolutions of ESHT were measured for both the NIR and IR transitions.[8b, 8f] No difference was found between the NIR and IR time-evolutions, thus the electron-proton decoupling does not occur in the smaller clusters, or it cannot be detectable by our time resolution of 3 ps. It is sharp contrast to the decoupling observed in the $n$ = 5 cluster.

To shed the light to the difference of the ESHT mechanism, the potential energy curves of the excited states are calculated. In the previous letter paper, the preliminary calculations assuming the structures of $C_s$ symmetry have been reported in Supporting Information.[15] However, the clusters have no molecular plane in $S_0$ thus we applied the excited state calculations under the $C_1$ symmetry. The initial geometry was taken from the previous report.[8d] Figure 8 shows the potential energy curves of the excited states in PhOH–$(NH_3)_n$ with A) $n$ = 3 and B) $n$ = 5. The potential energy curves are obtained by stretching the OH distance $R_{OH}$ keeping the other coordinate fixed at the geometry in $S_0$. This assumption means that the calculated potential curve is not the minimal energy path of the reaction. However, it is reasonable assumption to know the reaction mechanism because the electron and the proton is significantly lighter than overall clusters and their motion will be faster than the dynamics of overall geometrical change. On the other hand, we should not discuss the reaction mechanism after the complete proton/H atom transfer ($R_{OH}$ > ≈1.7 and ≈1.6 Å for $n$ = 3 and $n$ = 5, respectively) because the clusters release the reaction product •$H(NH_3)_n$ when ESHT completes. Such dynamics cannot be included in these potential curves, and thus the calculated energies increase at longest $R_{OH}$ because of the effect of the fixed geometries. It causes minima at ≈1.7 and 1.6 Å for $n$ = 3 and $n$ = 5, respectively, but these are not the real minima because of the dissociation of the products along the N-O coordinate. Thus, we neglect these minima in further discussion.

Both the $n$ = 3 and 5 clusters have $C_1$ symmetry and thus π/σ symmetry does not exist in principle but should be considered as labels. We attempt to distinguish the major character of the excited states from their shapes of orbitals. If the orbital is mainly localized on the aromatic ring like valence orbital, we call it π or π*. In the case that the orbital is diffuse likely to Rydberg orbital, it is denoted as σ*. For the $n$ = 3 cluster, the $S_1$ state in the region where $R_{OH}$ is shorter than CI has almost pure ππ* character and the mixing of the πσ* character is less than 1%. The $S_2$ state in the same region has high πσ* character (more than >74 %) and ππ* character is minor in this approximation. In contrast, ππ* and πσ* characters are significantly mixed in the $n$ = 5 cluster ($S_1$: >51 % of ππ*, $S_2$: >63 % of πσ*). For convenience, we also describe the excited states of $n$ = 5 as ππ* and πσ* according to the major character but the strongly mixed characters are the significantly different from those of $n$ = 3. The calculated energies of the ππ* and πσ* states are plotted by blue squares and red circles, respectively. In both clusters, the potential curves of ππ* and πσ* cross at ≈1.2 Å (diabatic representation). In the CC2 approximation, the avoided crossing on CI cannot be treated, thus the both potential curves are obtained before and after CI. In the real molecular clusters, these potentials should be repelled each other and will make the new potential curves, which are indicated by solid curves (adiabatic representation). The O-N distance in $S_0$ is 2.74 Å and 2.57 Å, thus the $R_{OH}$ ≈ 1.7 Å and ≈ 1.6 Å corresponds to the position that proton/H atom is transferred to the ammonia moiety in $n$ = 3 and 5, respectively. $R_{OH}$ in the ground state is ≈1.0 Å in both clusters, so that the ESHT reaction is thought to start from the points A ($n$ = 3) or C ($n$ = 5) in the Figure 8.

In the $n$ = 3 cluster, the ππ* state has the minimum at $R_{OH}$ ≈ 1.0 Å while the minimum of the πσ* state is $R_{OH}$ ≈ 1.7 Å. As a result, the avoided crossing of ππ* and πσ* produces the potential curve of $S_1$ (black solid curve) with a barrier at CI. The photoexcitation prepares the cluster at around the local minimum in $S_1$ by the Franck-Condon principle. Then the proton has to go through the barrier by tunneling to complete the photochemistry as in smaller complex and a free PhOH.[12] It is consistent to the relatively slow ESHT reaction of 24 ps lifetime, which are measured by both the NIR and IR time-resolved spectroscopy.[8b, 8f] After the tunneling, the character of the $S_1$ electronic state changes to the $\pi\sigma^*$ state in which the electron is transferred to the diffuse Rydberg-like orbital, and the NIR absorption at 8000 $cm^{-1}$ appears due to the 4p-$\sigma^*$ transition. Thus the lifetime of 24 ps measured by the TRNIR spectroscopy corresponds to the time required for the tunneling.[8b, 8f] The IR absorption of the NH stretching vibration also arises with the same time constant in the $n$ = 3 cluster.[8b, 8f] Then the NH bond formation must be complete within our experimental time-resolution after the tunneling. It means that the proton movement to the most stable position and dissociation to the product $\bullet H(NH_3)_3$ have to be finished within 3 ps.

The potential curves of the $n$ = 5 cluster are significantly different from those of the $n$ = 3 cluster. The $\pi\pi^*$ state (blue squares in Figure 8b) has the global minimum at the long $R_{OH}$ (≈1.5 Å, indicated by a letter D), which corresponds to the proton transferred structure. Although the potential curve of $\pi\sigma^*$ (red circles) is similar to that in $n$ = 3, the different shape of the $\pi\pi^*$ state causes the significant difference in the potential curves of $S_1$ (black curve) and $S_2$ (gray curve). The potential curve of $S_1$ does not have a barrier and smoothly go down over CI. Then the reaction takes place without barrier from the Franck-Condon position (≈1.0 Å, indicated by a letter C). It gives the fast motion to the proton along the reaction coordinate. Another important difference is the strong $\pi\pi^*/\pi\sigma^*$ mixing of $S_1$ and $S_2$ in the $n$ = 5 cluster as mentioned above. It means that the NIR transitions at ≈8000 $cm^{-1}$ will be possible if the cluster is simply excited to $S_1$ because of its $\sigma^*$ component. It is consistent to the fast appearance of the NIR transition at 8000 $cm^{-1}$ within 3 ps, which corresponds to the electron transfer. When the proton moves over CI along the potential curve of $S_1$, the electronic character is changed to $\pi\sigma^*$ (blue square). Since $\pi\sigma^*$ is also mixed with $\pi\pi^*$ heavily in $n$ = 5, the amount of $\sigma^*$ character in $S_1$ is not change significantly by going over CI. Thus the time evolution of 8000 $cm^{-1}$ which indicate the $\sigma^*$ character will not be sensitive to CI, and the decay of the NIR transition at 8000 $cm^{-1}$ have to be induced mainly by the formation of the $T_d$ reaction product after the dissociation of the ammonia moiety. However, the fast proton motion caused by the barrierless reaction path is not straightforward to understand why the formation of the final reaction product $T_d$ takes ≈20 ps, which is observed in the ps-TRIR spectroscopy. The proton can reach to the most stable position (indicated by a letter E) of $\pi\sigma^*$ immediately from the FC region. In the $n$ = 3 cluster, the dynamics after the tunneling is within 3 ps. If we take the analogy to the $n$ = 3 cluster, the formation of $T_d$ should also be very fast, probably within 3 ps. Then the ≈20 ps lifetime of the $T_d$ formation is not consistent with this simple scheme and one should add another step to the reaction.

To solve this inconsistency, we would like to discuss the following model. Since the proton can move with significant speed from the initially prepared position in $S_1$, it will be able to go over IC and proceed to the $S_2$ state without following the $S_1$ potential curve (the dynamics occurs only the $\pi\pi^*$ state that is $S_1$ at short distance and $S_2$ after CI). This corresponds to the classical ESPT reaction. Then the proton can be trapped in $S_2$ for a while, and due to the Coulomb attraction this cluster cannot dissociate. The $S_2$ state relaxes by internal conversion to $S_1$ (that is the $\pi\sigma^*$ state), which immediately causes the dissociation and successive formation of the final $T_d$ product. In this scenario, the observed ≈20 ps lifetime of the formation of the $T_d$ product corresponds to the rate of the internal conversion, i.e. the lifetime of $S_2$. The $S_2$ state beyond CI is denoted as ππ*, however this state is also the strong mixture of ππ* and πσ*. Thus the NIR absorption at 8000 $cm^{-1}$ can appear even from $S_2$ beyond CI. The rise of the NH stretching vibration will corresponds to the internal conversion. Thus, this model fits to the observed time-evolutions. The last point to consider is whether the cluster trapped in $S_2$ can be detected or not. In our observation, there is no evidence for the formation of $S_2$. Since the cluster in $S_2$ should have characteristic vibrational spectrum, we must discuss why the ps-TRIR spectrum is insensitive to such species.

Theoretical IR spectra are calculated at several points of potential curves for $n$ = 3 and 5, which are indicated by letters A–E in Figure 8. The molecular orbitals at these points are shown in Figure 10. The calculated IR spectra are shown in Figure 9. The calculated points A–E are shown together with theoretical spectra as A)–E), respectively. A) and C) show IR spectra right after the photoexcitation. The IR spectrum at D) corresponds to the one in $S_2$, which have the proton transferred structures of $n$ = 5. This spectrum is obtained without diffuse functions (in absence of diffuse functions, the electronic wavefunction of the πσ* state is poorly described and thus it becomes higher in energy than the ππ* state and this allows the normal mode analysis on the lowest excited state). The spectra in Figure 9 B) and E) are the IR spectra of the H transferred structure of the $n$ = 3 and 5 clusters, respectively. The calculated spectra of •$H(NH_3)_n$ ($n$ = 3 and 5) are also shown in Figure 9 F) and G). The vibrations at around 3000 $cm^{-1}$ or lower are H-bonded OH or NH stretching vibrations. Because of the strong H-bonding, these vibrations are expected to be broadened and will not contribute to the sharp bands in the ps-TRIR spectra. The bands at ≈3200 $cm^{-1}$ are free NH stretching vibrations and will be detected in the ps-TRIR spectra clearly. For $n$ = 5, the reaction proceeds from C to D and arrive E where the reaction product is generated immediately. The spectrum C at the initially prepared FC region does not have strong vibrational bands. The electronic state contains significant amount of πσ* orbital, which can enhance the oscillator strength of vibration by its diffuse σ* orbital. It is because vibrational motions in the diffuse σ* electron cloud easily change the dipole moment. However, at the FC region, the diffuse σ* is rather distributed outside of the cluster and does not affected by the vibrations (See Figure 10 C). The theoretical IR spectrum at D corresponding to the "hidden" $S_2$ state has slightly stronger free NH vibrations than the FC region, however the enhancement is limited to 2–3 times. On the other hand, the spectrum in Figure 9 E) shows more than around ten times stronger vibrational transitions than those in the FC region (Figure 9 C)). Here, the σ* orbital covers on •$H(NH_3)_5$ (Figure 10 E) thus all the NH stretching vibrations are significantly enhanced. From the point E, the cluster will release the $T_d$ product •$H(NH_3)_5$ immediately. The $T_d$ product has 20 times stronger vibrational intensity than the $S_2$ cluster by the complete overlap of σ* and the NH bonds (Figure 10 G). Thus, from the vibrational calculations, we can understand that the $S_2$ state cannot be detected because of its weak vibrational intensity.

The theoretical IR spectra for $n$ = 3 give essentially the same explanation for the enhancement of the IR transitions at the destination of the reaction (See Figure 9 A), B), and F)). The σ* orbitals are also shown in Figure 10 B and F. The difference from $n$ = 5 is that the vibrations are not enhanced at the bottom of $S_1$ (point B) where the H transfer is completed inside of the cluster (see Figure 9 B)). The weak enhancement at B is little overlap of the σ* orbital to the ammonia moiety in the $n$ = 3 cluster (see Figure 10 B). The strong enhancement of the NH vibrations in the product •$H(NH_3)_3$ is the same as the product of the $n$ = 5 cluster, and the formation of the reaction product should be fast enough. Thus the vibrational enhancement by the product formation is also reproduced in the theoretical calculations.

**4. Conclusion**

We have measured the ps-TRNIR and ps-TRIR spectra of photoexcited PhOH–$(NH_3)_5$, and traced the electron transfer and proton motion separately by the Rydberg type NIR transitions and the NH vibrational transitions. The electron transfer monitored by the NIR transition rises within 3 ps while the overall H transfer detected by the IR absorption of NH vibrations appears with ≈20 ps lifetime. The different time-evolution clearly proves that the electron transfer and proton movement is decoupled in ESHT of PhOH–$(NH_3)_5$. On the other hand, the previous experiment on PhOH–$(NH_3)_3$ does not show the difference of the time-evolutions between the NIR absorption and the IR transitions.[8b, 8f] The excited state calculations on both clusters give significant difference in the potential energy curves in the $S_1$ and $S_2$ states. The potential curve of the $n$ = 3 cluster has a potential barrier from the Franck-Condon region from $S_0$ along the OH bond length while that of the $n$ = 5 cluster does not have. It explains that the ESHT rate in the $n$ = 3 cluster is slower than that in $n$ = 5 due to the proton tunneling. This difference arises from the stability of the proton transferred structures. The proton transferred structure is more stable than the FC region in $n$ = 5, while it is not energetically accessible in $n$ = 3. It should be noted that the proton transferred structure in PhOH–$(NH_3)_n$ clusters have been experimentally explored in long time,[3b, 3d, 4b-d, 6a] but the reason why it could not be detected is that the H transferred structure is more stable and the proton transferred structure is quenched by the internal conversion. Instead, the proton transferred structure is contributed to trap the reactive cluster for a while, and causes the delayed appearance of the IR absorption corresponding to the complete hydrogen transfer from the electron transfer. The weak IR transitions calculated for $S_2$ (the proton transferred structure) is consistent to the observed ps-TRIR spectra that do not show the signature of the trapped species. In both the $n$ = 3 and $n$ = 5 clusters, the excited state hydrogen transfer dynamics are well explained by the assumption that the electron transfer and the proton motion are separated. We believe this work provides the fundamental understanding how proton and electron are related in the H transfer reactions with experimental evidence and high level theoretical calculations. The description above is based on static calculations (potential energy curves) on one dimension only and most probably there will be very valuable information to test this description if one can perform dynamical calculations.

**Acknowledgements**

This work was supported in part by KAKENHI (JP205104008) on innovative area (2503), KAKENHI (JP15H02157, JP15K13620, JP16H06028), and the Cooperative Research Program of the "Network Joint Research Center for Materials and Devices" from the Ministry of Education, Culture, Sports, Science and Technology (MEXT), Japan. All the calculations were performed by the computing facility cluster GMPCS of the LUMAT federation (FR LUMAT 2764).

**Keywords**

Excited state hydrogen transfer • Solvated clusters • Proton coupled electron transfer • Phenol • πσ* state

**Entry for the Table of Contents**

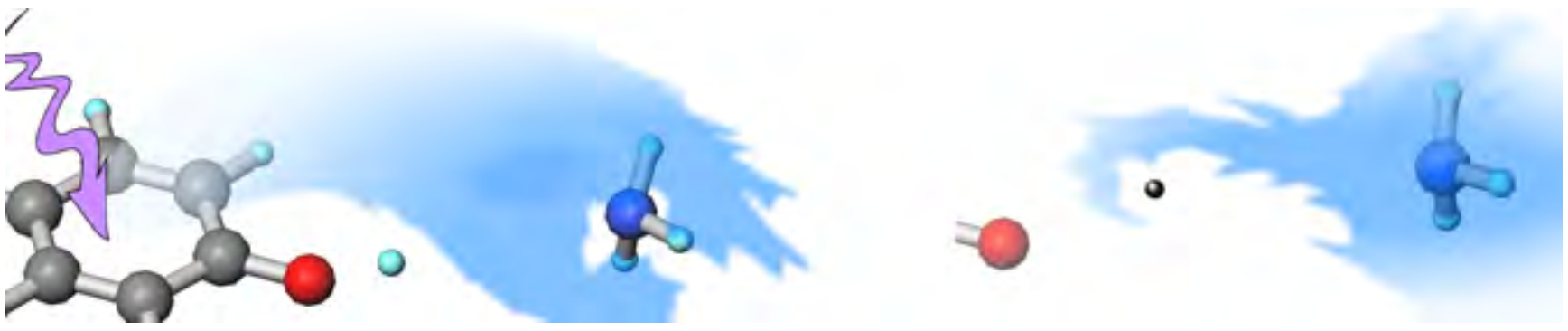

**Decoupling of the electron and proton of a hydrogen atom in the excited state hydrogen transfer (ESHT)**: Time-resolved NIR and IR spectroscopies coupled with ab initio calculation on solvated clusters of phenol by ammonia established that initial ultrafast delocalization of an electron over the solvents by photoexcitation next eliminates the proton from phenol to retrieve the hydrogen atom in the solvents moiety.

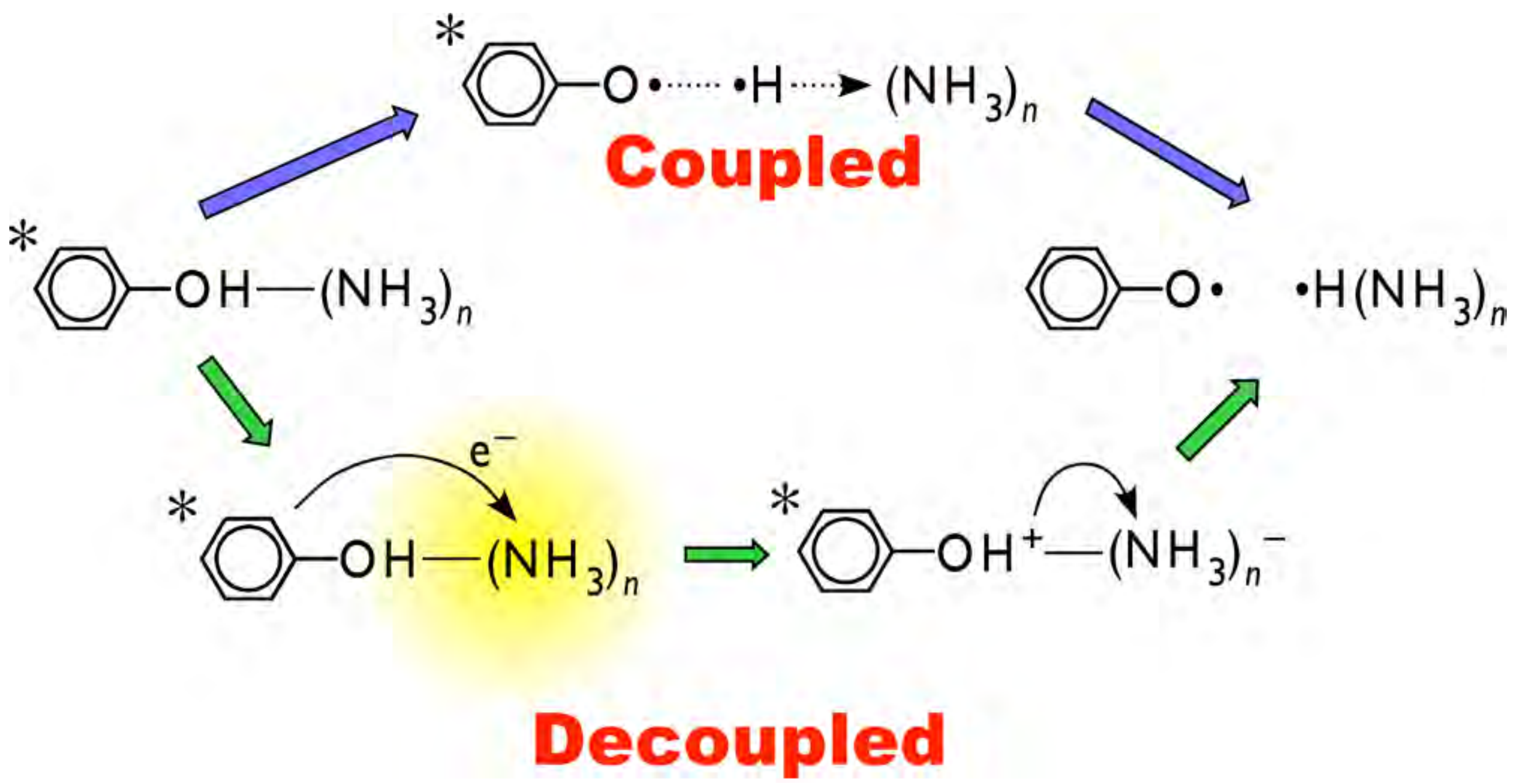

**Figure 1** Schematic description of two distinct reaction pathways of the excited state hydrogen transfer reaction of PhOH–$(NH_3)_n$ clusters. Upper path, coupled mechanism, where the electron and proton transfer together from phenol to the ammonia moiety keeping status of a hydrogen atom. Lower path, decoupled mechanism, where the electron and the proton independently move to the ammonia moiety. Resultant radical products are the same in both the cases.

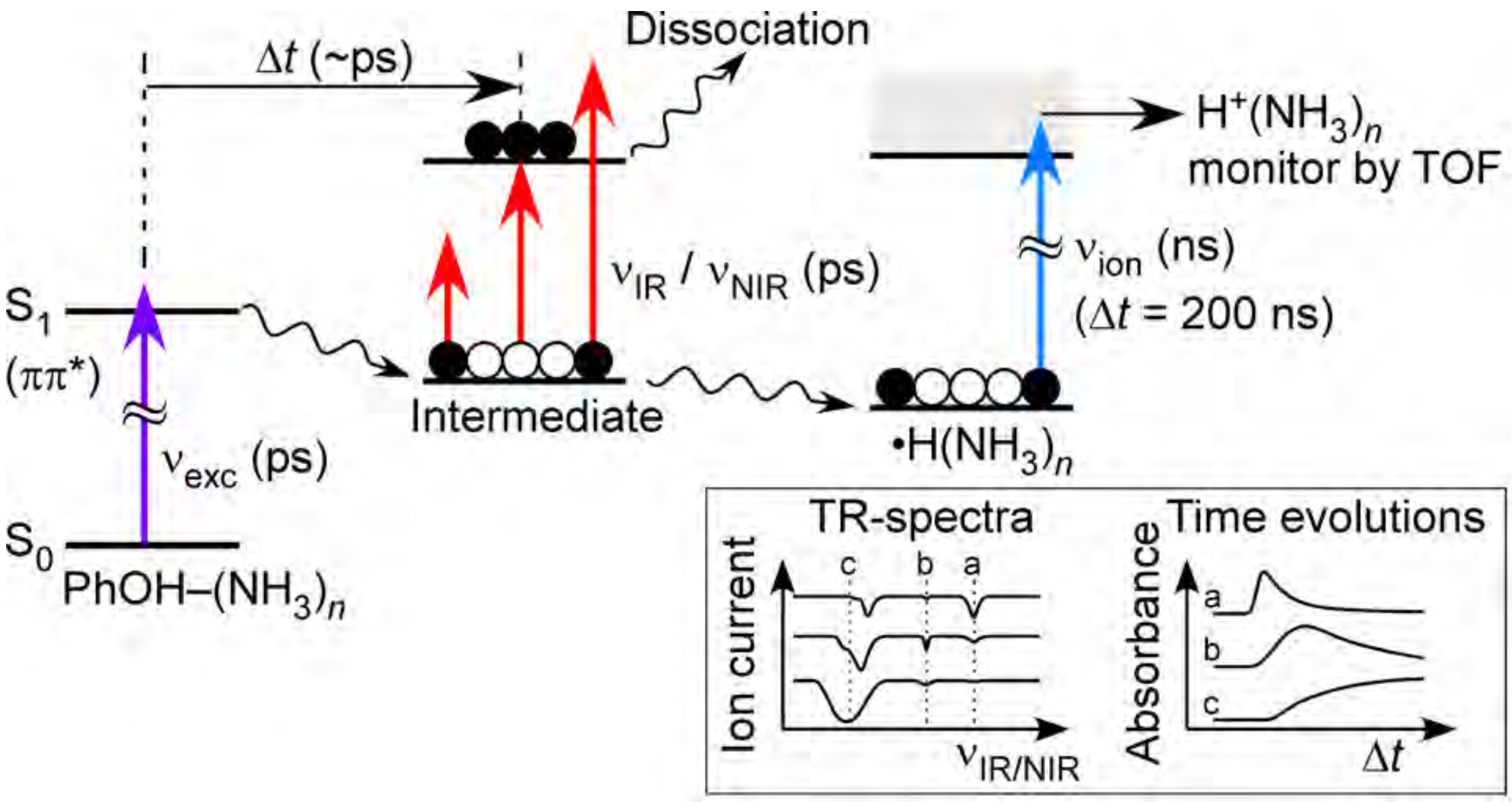

**Figure 2** Schematic diagram of the principle of picosecond IR/NIR spectroscopy of PhOH–$(NH_3)_n$ clusters.

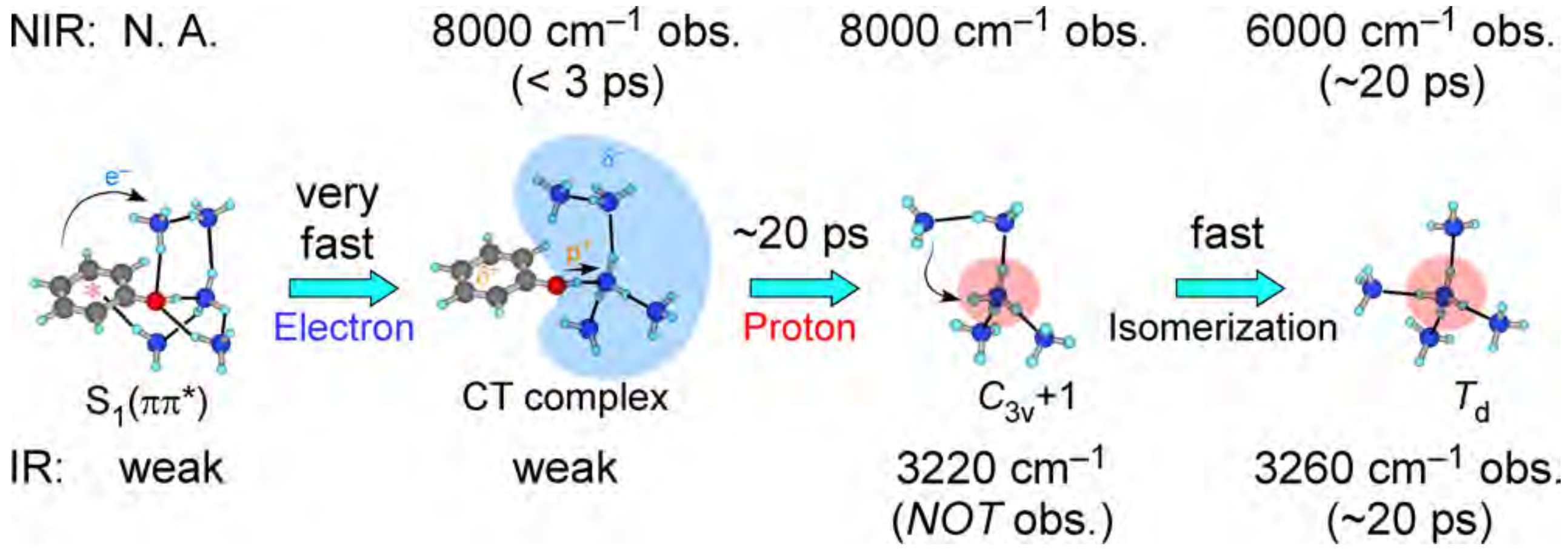

**Figure 3** Structures and responsible absorption features (upper side: electronic transitions in the NIR region, lower side: NH stretching transitions in the 3 μm region) of the PhOH–$(NH_3)_5$ cluster and the •$H(NH_3)_5$ product along the course of the decoupled ESHT reaction mechanism.

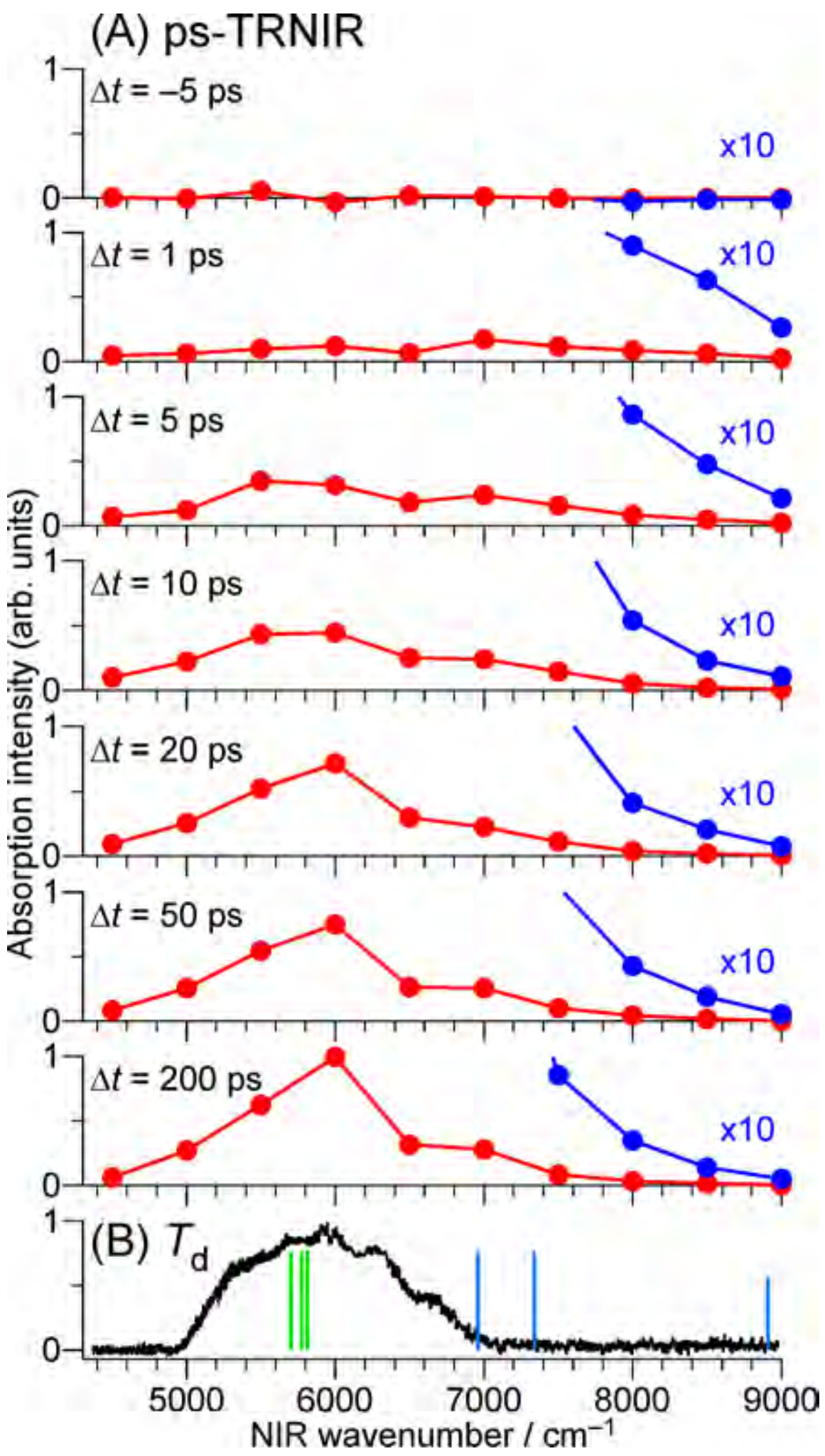

**Figure 4** (A) Picosecond time-resolved NIR spectra of PhOH–$(NH_3)_5$ from $\Delta t$ = −5 to 200 ps. (B) The NIR spectrum measured by nanosecond lasers at $\Delta t$ = 180 ns. Green and blue bars represent theoretical spectra of the $T_d$ and $C_{3v}$+1 isomers, respectively. A common scale is used for vertical axes of time-resolved spectra to compare absorption intensities in different delay times.

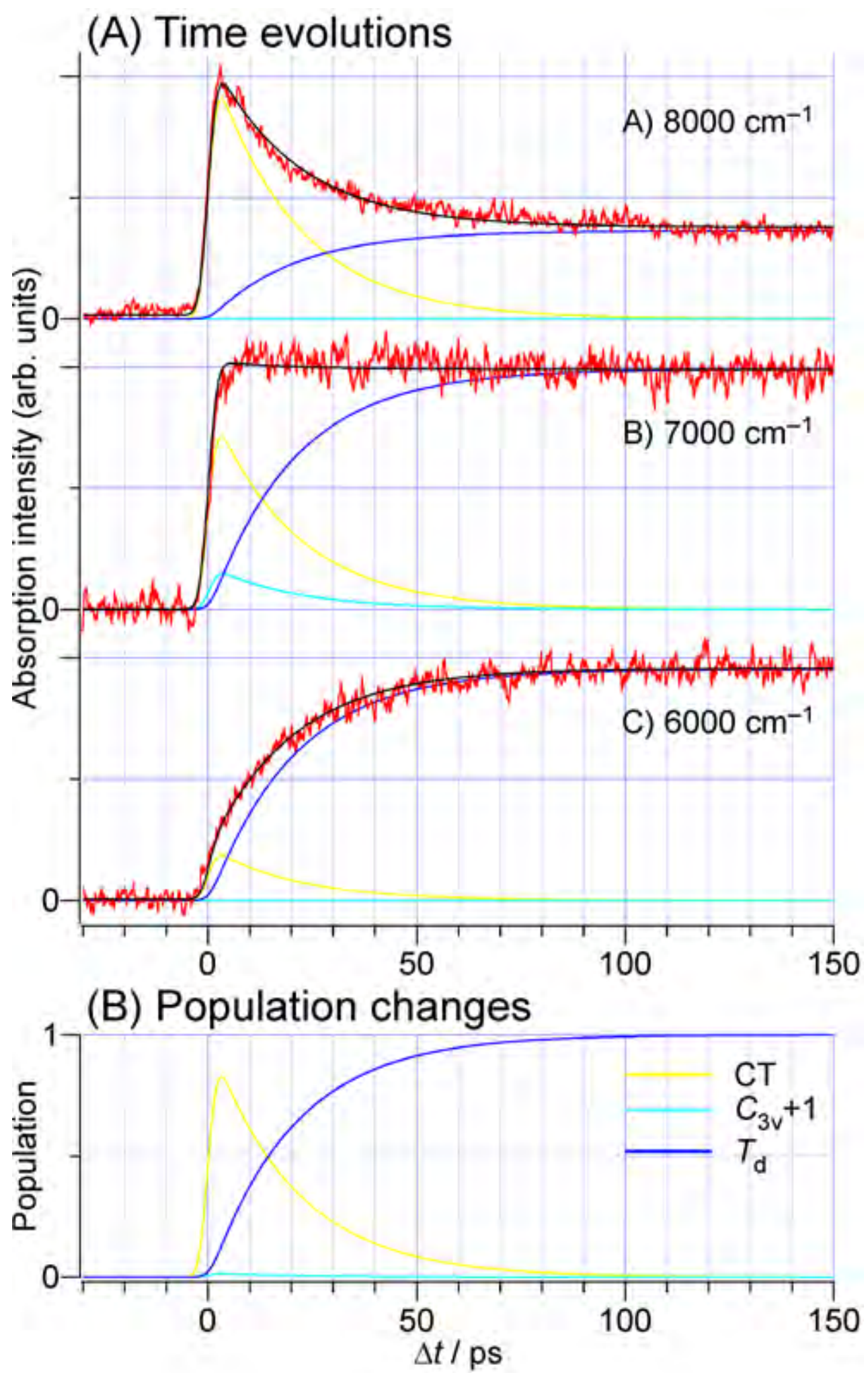

**Figure 5** (A) Time evolution of the transient NIR absorption of PhOH–$(NH_3)_5$ after the $\pi\pi^*$ excitation probed at A) 8000 $cm^{-1}$, B) 7000 $cm^{-1}$, and C) 6000 $cm^{-1}$. Black curves show the best fitted curves obtained by the global fitting, see text. Curves in yellow, light blue, and blue represent decomposition of the contribution from CT, $C_{3v}$+1, and $T_d$ species, respectively. (B) Population changes of these species derived by the fitting results. The total population is normalized to one. Small population of $C_{3v}$+1 isomer reflects its short lifetime.

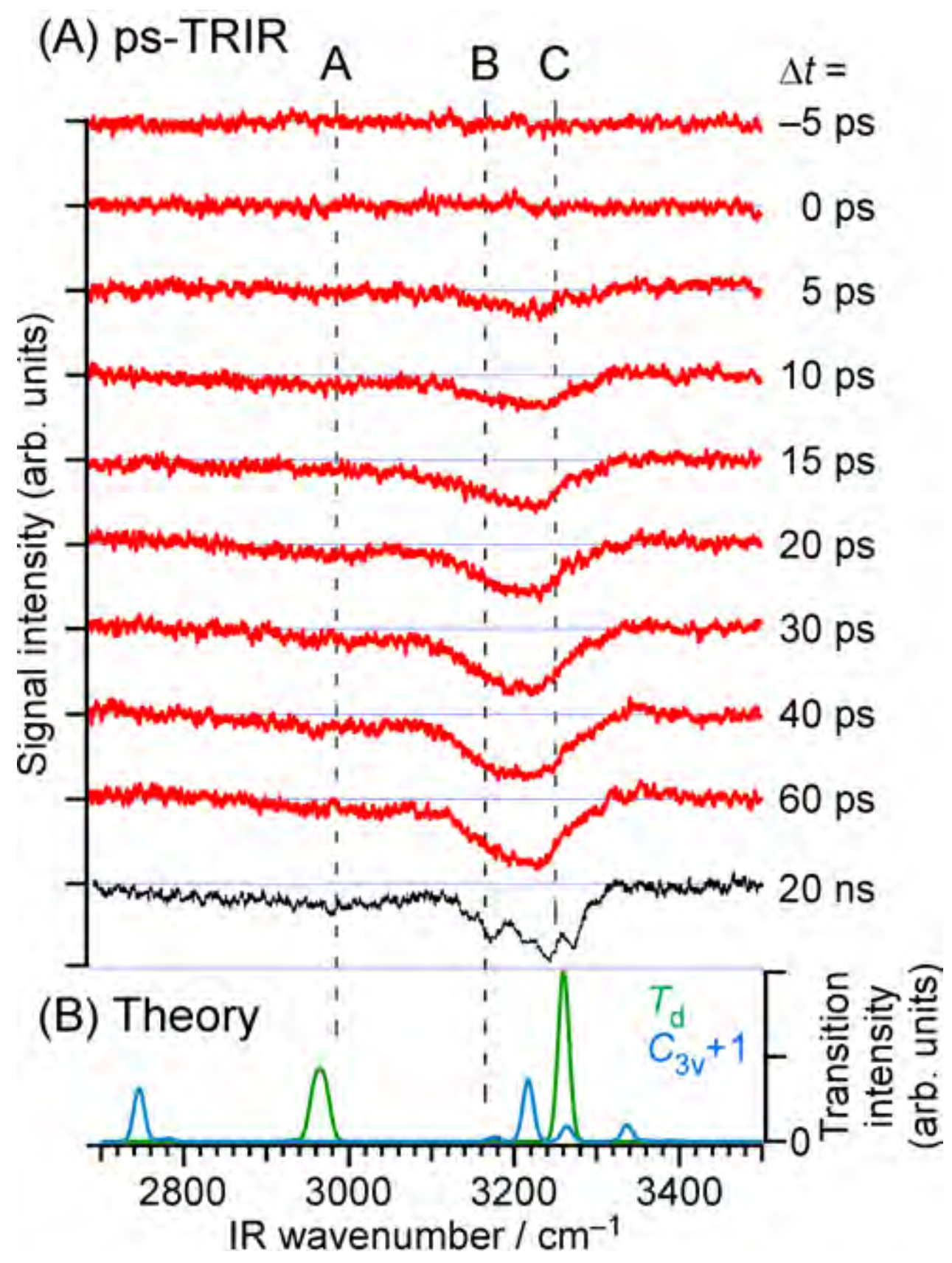

**Figure 6** (A) Time-resolved IR spectra in the NH stretching vibrational region of PhOH–$(NH_3)_5$ after the $\pi\pi^*$ excitation. The spectrum at $\Delta t$ = 20 ns was measured using nanosecond lasers, and was adapted from previous data published in reference 48. (B) Theoretical spectra of $T_d$ and $C_{3v}$+1 isomers of the ESHT products, •$H(NH_3)_5$, are also shown at the bottom as green and blue traces, respectively. These theoretical traces are also plotted based on data in ref. 8f. Positions probed by the time evolution measurements shown in Figure 7 are indicated by broken lines labeled A, B, and C.

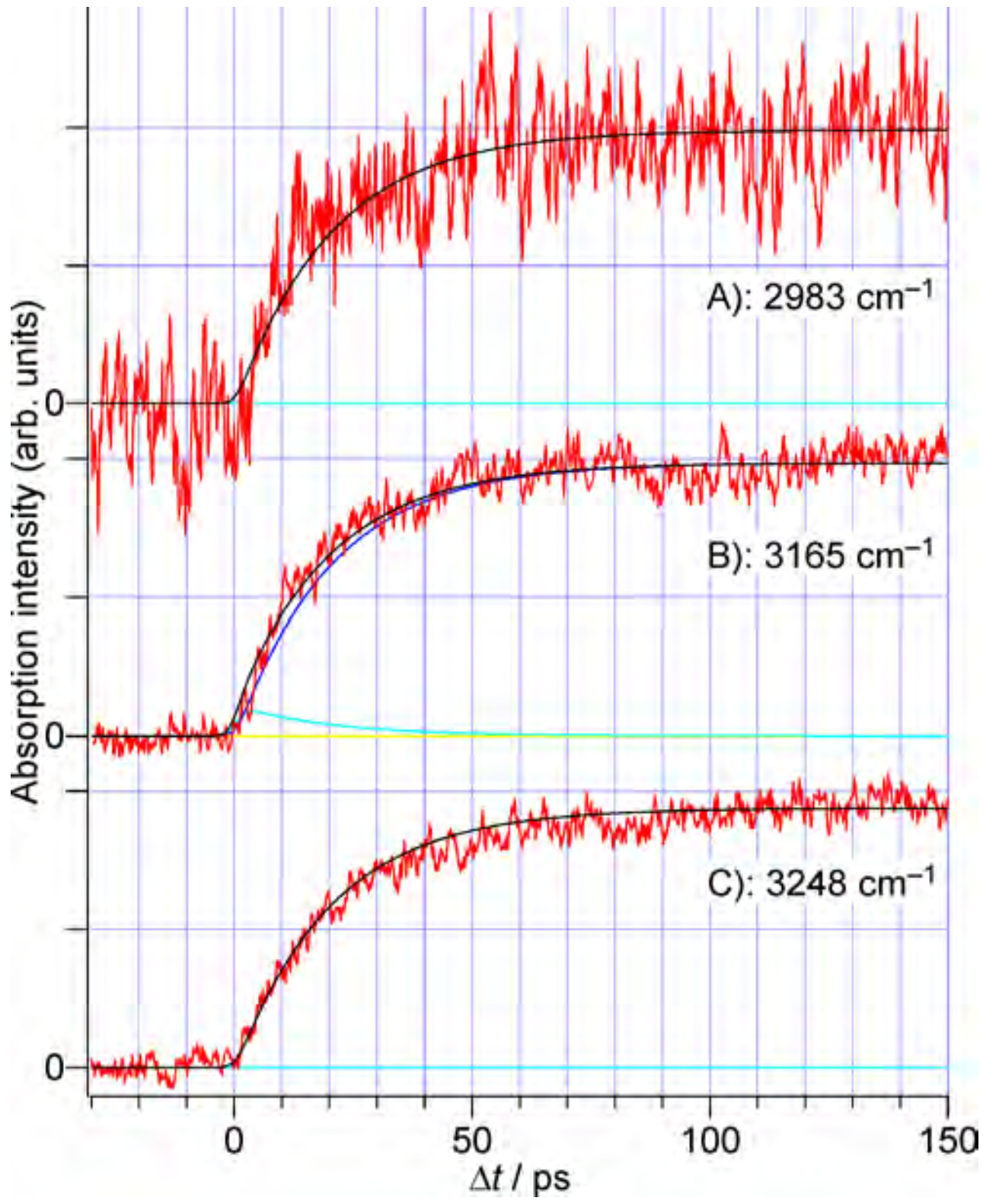

**Figure 7** Time evolutions of transient NH stretching vibrations of PhOH–$(NH_3)_5$ after the $\pi\pi^*$ excitation probed at A) 2983 $cm^{-1}$, B) 3165 $cm^{-1}$, and C) 3248 $cm^{-1}$, respectively. Best fitted curves and population decomposition are also presented with the same format as in Figure 5. In A) and C) the best fitted curves completely overlap on contribution from the $T_d$ isomer.

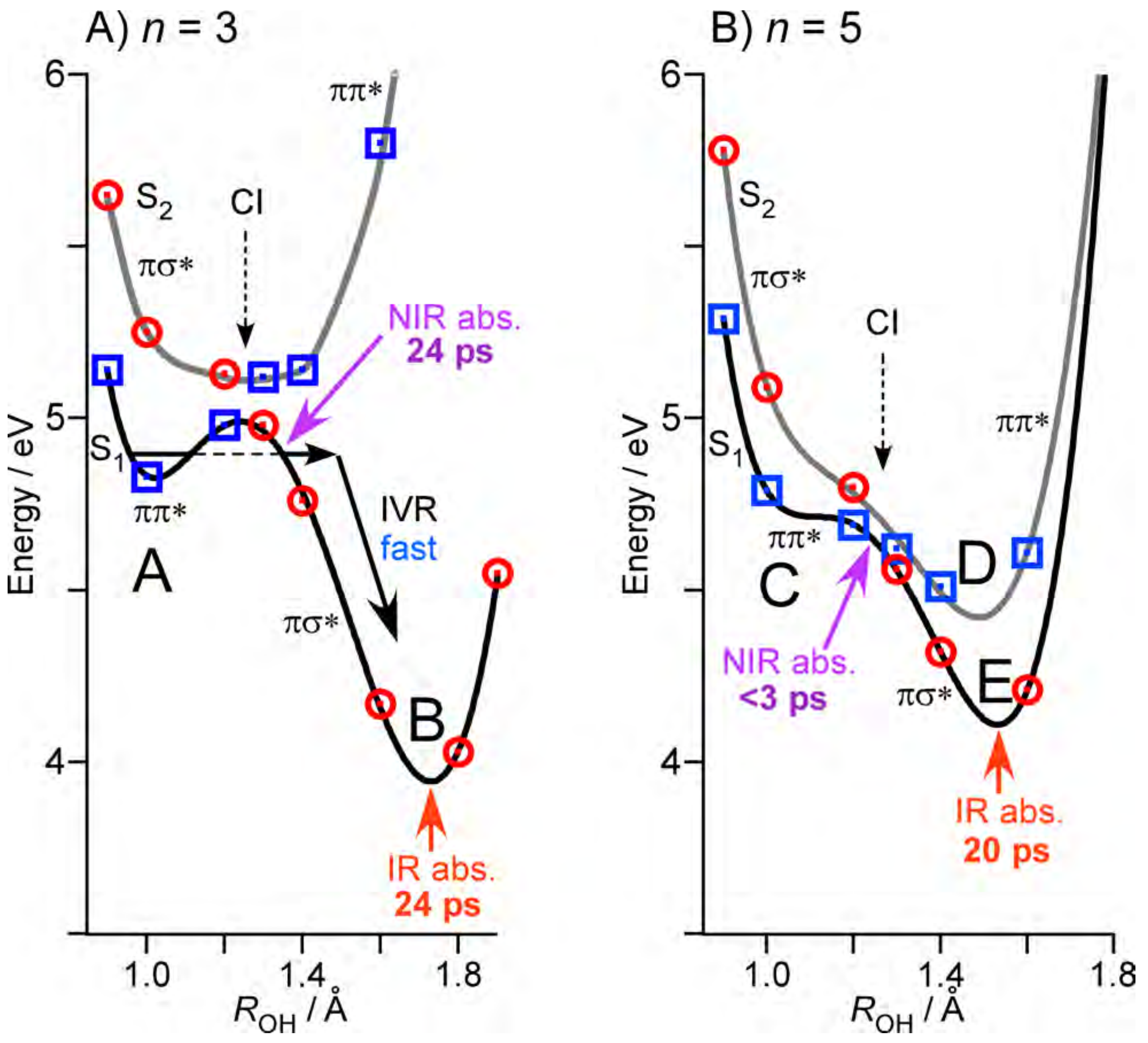

**Figure 8** Potential curves of ππ* and πσ* states of A) PhOH–$(NH_3)_3$ and B) PhOH–$(NH_3)_5$ obtained by CC2/aug-cc-pVDZ level of theory. Nature of the electronic state of the lowest curve, denoted as $S_1$, changes from ππ* to πσ across CI at $R_{OH} \approx 1.3$ Å, and vice versa for the second curve denoted as $S_2$. Labels A–E represent positions where theoretical IR spectra shown in Figure 9 and molecular orbitals shown in Figure 10 are calculated. Energy increase and resultant minima in regions $R_{OH} > 1.5$ Å are artificial due to approximations imposed by the calculations, see text.

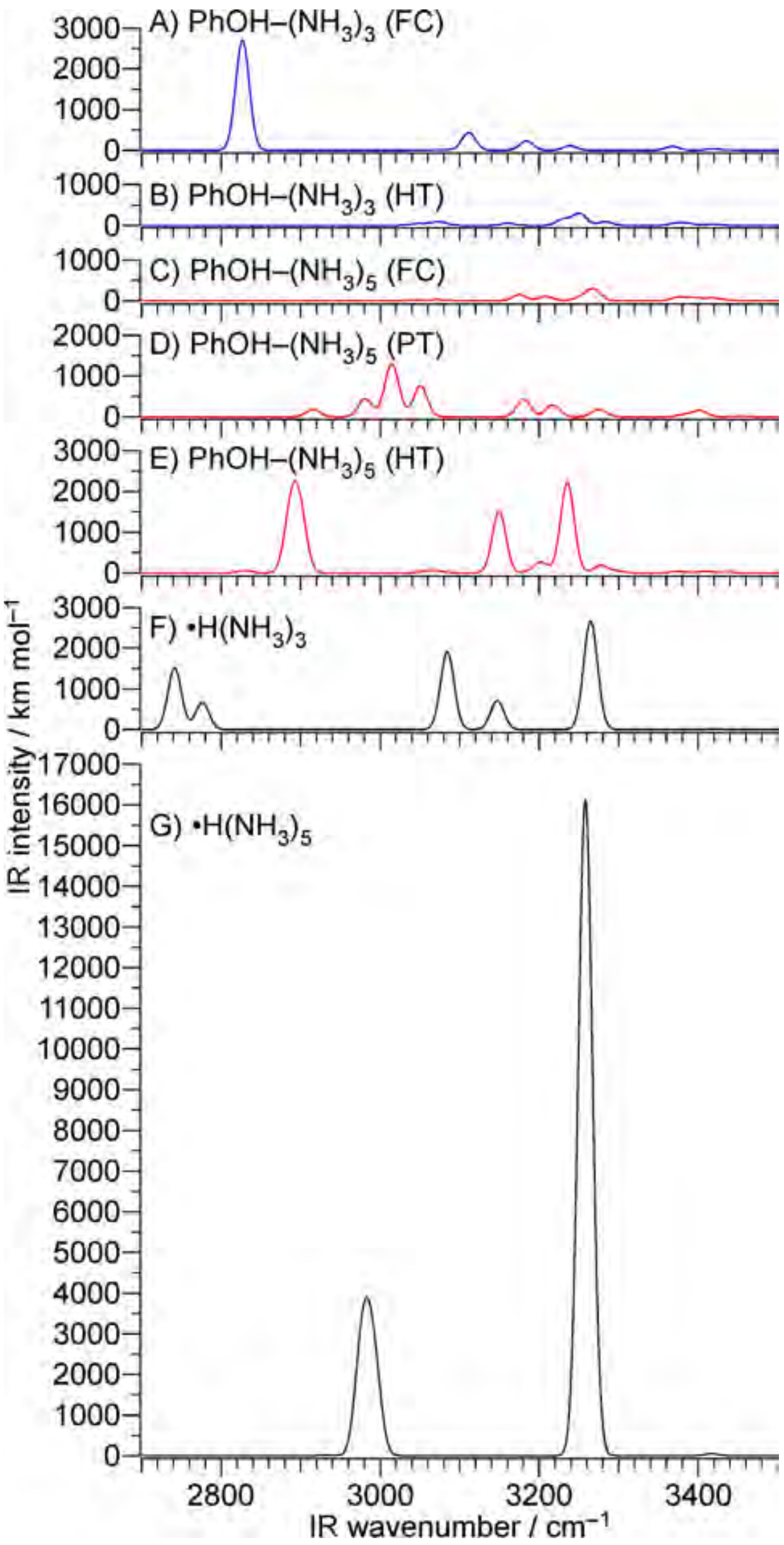

**Figure 9** A)–E) Theoretical IR spectra in the free NH stretching vibrational region of structures corresponding to A–E shown in Figure 8. F) and G) are IR spectra of the product radicals. A common vertical scale is used to all the spectra for the sake of comparison of the IR intensities.

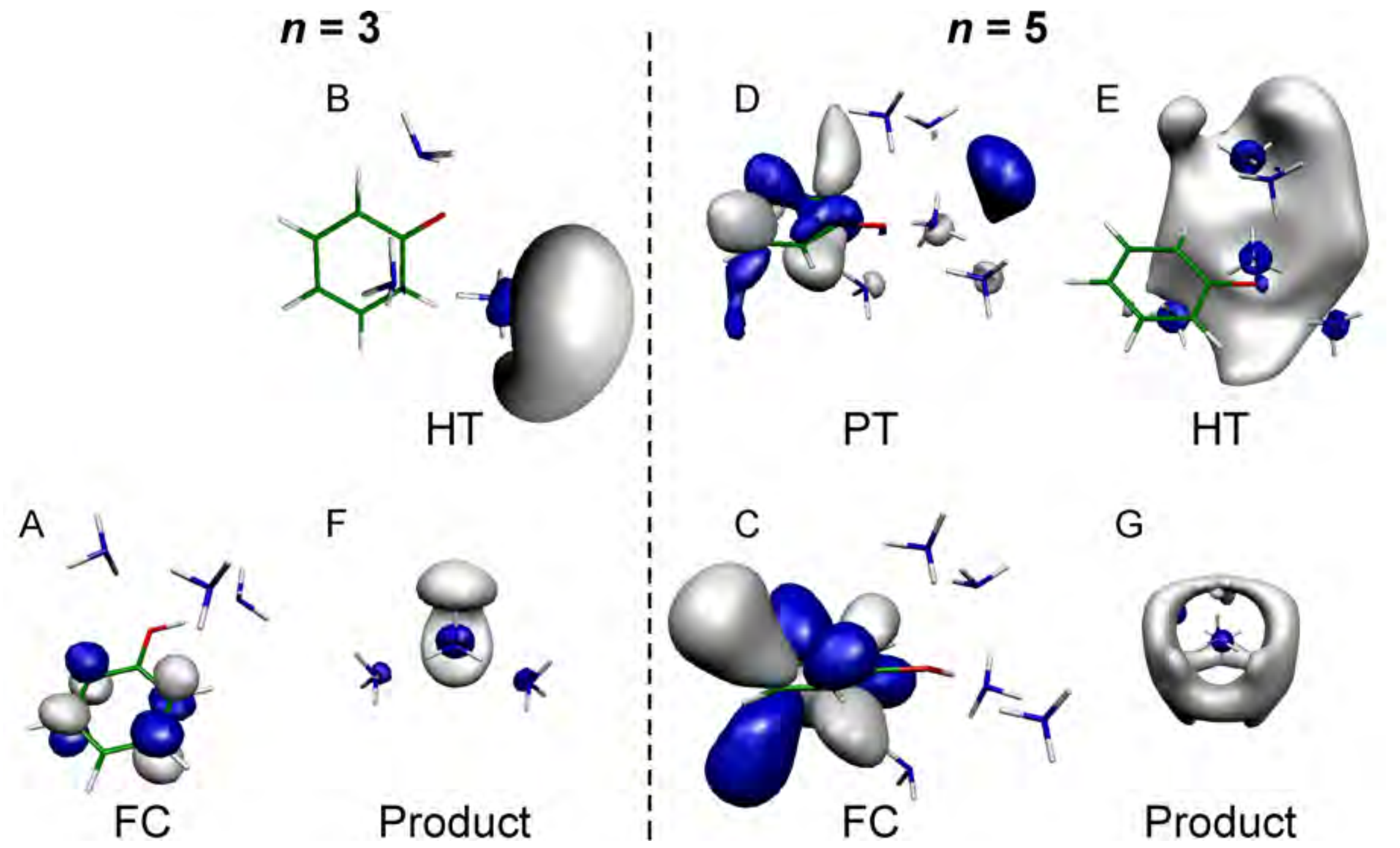

**Figure 10** Shapes of the LUMO orbital of structures corresponding to positions A–E in Figure 8. F and G show shapes of SOMO orbitals of the product radicals.

# Supporting Information

Electron-proton transfer mechanism of excited state hydrogen transfer in phenol–$(NH_3)_n$ ($n$ = 3 and 5)

by M. Miyazaki, R. Ohara, C. Dedonder, C. Jouvet, and M. Fujii

Contents:

(A) Experimental setup

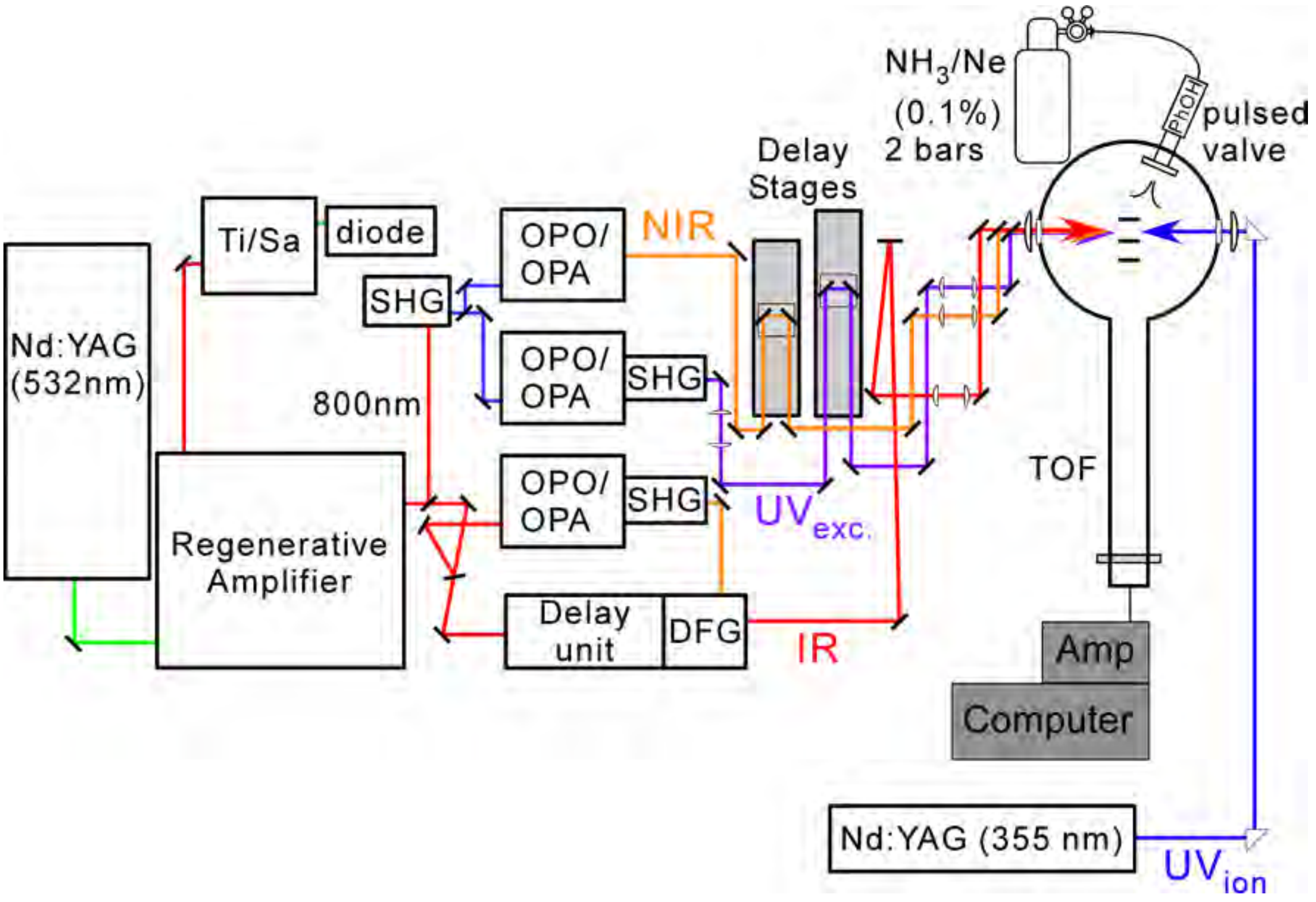

**Figure S1** A schematic diagram of the experimental setup of the picosecond time-resolved IR/NIR spectroscopy for PhOH–$(NH_3)_n$ clusters.

(B) Fitting procedure

In the fitting, a sequential two step reaction, A → B → C, was assumed for this reaction. Here, A corresponds to the CT complex, B to the $C_{3v}$+1 isomer, and C to the $T_d$ final product shown in Figure 3, respectively. The initially prepared ππ* state was not taken into account because conversion to the CT complex (i.e. the electron transfer rate, $k_{S_1-CT}$) is much faster than time resolution of our picosecond laser system, as can be seen from the fast rise of NIR absorption probed at 8000 cm$^{-1}$. Thus, the CT complex can be recognized as the initial species of this reaction in the analysis.

Solutions of rate equations for this reaction scheme are represented by double exponential functions:

$$\frac{[\mathrm{CT}](t)}{[\mathrm{S_1}]_0} = \mathrm{e}^{-k_{\mathrm{CT}-C_{3v}+1}(t-t_0)}\,\mathrm{H}(t-t_0), \tag{1}$$

$$\frac{[C_{3\mathrm{v}}{+}1](t)}{[\mathrm{S}_1]_0} = \left[\mathrm{e}^{-k_{C_{3\mathrm{v}}+1-T_\mathrm{d}}(t-t_0)} - \mathrm{e}^{-k_{\mathrm{CT}-C_{3\mathrm{v}}+1}(t-t_0)}\right] \frac{k_{\mathrm{CT}-C_{3\mathrm{v}}+1}}{k_{\mathrm{CT}-C_{3\mathrm{v}}+1} - k_{C_{3\mathrm{v}}+1-T_\mathrm{d}}} \mathrm{H}(t-t_0), \tag{2}$$

$$\frac{[T_\mathrm{d}](t)}{[\mathrm{S}_1]_0} = \left[\left\{\left(1-\mathrm{e}^{-k_{C_{3\mathrm{v}}+1-T_\mathrm{d}}(t-t_0)}\right)\right\} k_{\mathrm{CT}-C_{3\mathrm{v}}+1} - \left\{1 - \mathrm{e}^{-k_{\mathrm{CT}-C_{3\mathrm{v}}+1}(t-t_0)}\right\} k_{C_{3\mathrm{v}}+1-T_\mathrm{d}}\right] \frac{1}{k_{\mathrm{CT}-C_{3\mathrm{v}}+1} - k_{C_{3\mathrm{v}}+1-T_\mathrm{d}}} \mathrm{H}(t-t_0), \tag{3}$$

where $[\mathrm{X}]/[\mathrm{S}_1]_0$ and $\mathrm{H}(t-t_0)$ represent relative population of species X and the unit step function, respectively, and $k_{\mathrm{CT}-C_{3\mathrm{v}}+1}$ and $k_{C_{3\mathrm{v}}+1-T_\mathrm{d}}$ are rate constants of the first and the second steps. Time zero of the reaction, $t_0$, was also treated as a fitting parameter.

The instrumental function was considered by convolution of a Gaussian function:

$$\int_{-\infty}^{\infty} G(T-t)\,\mathrm{e}^{-kT}\,\mathrm{H}(T)\,\mathrm{d}T = \frac{1}{2}\,\mathrm{e}^{-kt+\frac{\sigma^2k^2}{2}}\left\{1+\mathrm{erf}\left(\frac{t}{\sqrt{2}\,\sigma}-\frac{\sigma k}{\sqrt{2}}\right)\right\}, \tag{4}$$

where $G(t) = \frac{1}{\sigma\sqrt{2\pi}}\mathrm{e}^{-\frac{t^2}{2\sigma^2}}$ with $\sigma = \frac{FWHM}{2\sqrt{2\ln 2}}$, and *FWHM* represents width of the instrumental function, which is typically ca. 3 ps for our system.

Since the species considered here have rather broad and strong absorption features, particularly in the NIR region, overlapping of the absorption should be taken account. In this analysis, we assumed that all the species contribute to whole the NIR region measured in this work, 5000–9000 $\mathrm{cm}^{-1}$. Thus, time evolutions of the NIR absorption were represented by linear combinations of population changes, (1)–(3).

For the NH stretching vibrations in the 3 μm region, contributions from the CT complex were excluded due to its low IR intensities compared to those of $C_{3\mathrm{v}}$+1 and $T_\mathrm{d}$ product radicals. According to the calculated band positions shown in Figure 6 (B), we assumed that at the position A (2983 $\mathrm{cm}^{-1}$) and C (3248 $\mathrm{cm}^{-1}$) the intensities exclusively originate from the $T_\mathrm{d}$ final product while at the position B (3165 $\mathrm{cm}^{-1}$) the absorption is composed of contributions from both the $C_{3\mathrm{v}}$+1 and $T_\mathrm{d}$ products.

A global fitting was performed considering all the measured time evolutions (three from the NIR region and three from the 3 μm region) under conditions mentioned above. The best fitted curve and population decomposition evaluated from the obtained fitting parameters and (convoluted) equations (1)–(3) are shown in Figure 5 and 7.